\documentclass{article}
\usepackage[dvipdfmx]{graphicx}
\usepackage{mediabb}
\usepackage{amsmath}
\def\be{\begin{equation}} 
\def\ee{\end{equation}} 
\def\bq{\begin{equation}} 
\def\eq{\end{equation}} 
\def\bqa{\begin{eqnarray}} 
\def\eqa{\end{eqnarray}}

\def\nle{\mathrel{\vcenter
     {\hbox{$<$}\nointerlineskip\hbox{$\sim$}}}}
\def\nge{\mathrel{\vcenter
     {\hbox{$>$}\nointerlineskip\hbox{$\sim$}}}}

\begin{document}


\begin{center}
{\large \bf One loop effects of natural SUSY in third generation fermion production at the ILC}

\end{center}
\vspace{0.2 cm}
\begin{center}

Yusaku Kouda$^{1,*}$, Tadashi Kon$^1$,  Yoshimasa Kurihara$^2$, Tadashi Ishikawa$^2$, \\ 
Masato Jimbo$^3$, Kiyoshi Kato$^4$ and Masaaki Kuroda$^5$\\
\vspace{4 mm}
$~^{1}~$ Seikei University, Musashino, Tokyo 180-8633, Japan\\
$~^{*}~$ E-mail: {dd146101@cc.seikei.ac.jp} \\
$~^{2}~$ KEK, Tsukuba, Ibaraki 305-0801, Japan\\
$~^{3}~$ Chiba University of Commerce, Ichikawa, Chiba 272-0827, Japan\\
$~^{4}~$ Kogakuin University, Shinjuku, Tokyo 163-8677, Japan\\
$~^{5}~$ Meiji Gakuin University, Yokohama, Kanagawa 244-8539, Japan\\
\end{center}

\noindent{\bf Abstract}\\
{
 Within the framework of the Minimal Supersymmetric Standard Model, we investigate the 1-loop effects of supersymmetric particles on the third-generation fermion-pair production at the ILC. Three sets of the SUSY parameters are proposed which are consistent with the observed Higgs mass, the muon $g$-$2$, the Dark Matter abundance and the decay branching ratios of B meson.  We discuss  the possibility of discovering the signals consistent with SUSY as well as of experimentally distinguishing the proposed sets of SUSY parameters.
}

\vspace{0.5cm}

\section{Introduction}

The Large Hadron Collider (LHC) has completed its first run and succeeded spectacularly in discovering the Higgs particle with a mass about $126$ GeV\cite{higgsatlus,higgscms}. This discovery has marked a landslide victory of the standard model (SM). 
In order to explain the elementary scalar Higgs with a mass $\mathcal{O}(10^2)$ GeV in the framework of the Grand Unified Theory (GUT),  we consider the minimal supersymmetric (SUSY) extension of the SM (MSSM)\cite{martin}. If there are SUSY particles (sparticles) with masses $10^2\sim 10^3$GeV, 
the reason why the mass of the Higgs particle is much smaller than the GUT scale is naturally understood. 
Moreover, the lightest SUSY particle (LSP) is a natural candidate for the dark matter (DM). 
We call such kind of theory ``natural SUSY".

First, we propose three sets of the SUSY parameters which are consistent with the experimental results of  (i) the Higgs mass, (ii) the muon $g$-$2$, (iii) the DM abundance, (iv) ${\rm{Br}}(b\to s\gamma)$, (v) ${\rm{Br}}(B_s\to \mu^+\mu^-)$ and (vi) the direct search for the sparticles  at the high energy colliders.  
Previous works often have only considered  either the DM abundance\cite{dmmssm1,dmmssm2,hamaguchi1} or the muon $g$-$2$\cite{wang,hamaguchi2} constraints on the MSSM. We assume that both  constraints  are simultaneously significant. 
We used SuSpect2\cite{suspect2}, SUSY-HIT\cite{susyhit} and micrOMEGAs\cite{micromega} to calculate the MSSM predictions for (i) $\sim$ (v). The next purpose of this paper is to study the  1-loop effects of sparticles in  the processes 
$e^-e^+$ $\to\tau^-\tau^+$, ${\bar{b}}b$, ${\bar{t}}t$ at the International Linear Collider (ILC)\cite{ilcd} for the selected parameters. For this purpose, we have used the automatic computation system $\tt GRACE$ for the MSSM\cite{grace1,grace2}.
We can expect the viable loop effects of even heavy sparticles with masses larger than the beam energy. 
 We have confirmed that the magnitudes of cross sections are consistent with  those presented in the previous work\cite{hollik}.

The paper is organized as follows. In Section 2, we discuss the selection of the values of MSSM parameters.  In Section 3 we
 show the numerical results for cross sections of the production processes
 at the ILC.  In Section 4, we give the summary and conclusions.


\section{Selection of the MSSM parameter sets}
\subsection{The muon $g$-$2$ anomaly}

There is a $3.5\sigma$ deviation between the SM prediction $a^{SM}_\mu$ and the experimental value $a^{exp}_\mu$ of the muon magnetic moment $g$-$2$\cite{g-2exp}, where 
$a_\mu=(g-2)/2$,
\begin{eqnarray}
a^{exp}_\mu&=&(1165920.91\pm0.54\pm0.33)\times{10}^{-9}, \\
a^{SM}_\mu&=&(1165918.03\pm0.01\pm0.42\pm0.26)\times{10}^{-9}.
\label{g2sm}
\end{eqnarray}
We consider that the deviation,
\begin{equation}
\Delta a_{\mu}=(2.88\pm0.63\pm0.49) \times{10}^{-9}
\end{equation}
comes from the MSSM contributions. Then it becomes a constraint on the MSSM parameters. 
We naively consider the theoretical uncertainty \footnote{The possible larger  error in the estimation of the higher order SM contributions are taken into account in Ref.\cite{ag2}.} of the SM prediction in eq.(2). 
The analytic expressions for the MSSM contributions to the muon $g$-$2$\cite{g-2} are given by
\begin{eqnarray}
a_\mu({\tilde W}-{\tilde H},{{\tilde \nu}_\mu})&=&\frac{g^2}{{8{\pi}}^2}\frac{{m_\mu^2{}M_2}{\mu}\tan{\beta}}{{m^4_{\tilde \nu}}}F_a\left(\frac{M_2^2}{m_{\tilde \nu}^2},\frac{\mu^2}{m_{\tilde \nu}^2}\right), \\
a_\mu({\tilde B,}{\tilde \mu_L}-{\tilde \mu_R})&=&\frac{g_Y^2}{{8{\pi}}^2}\frac{{m_\mu^2}{\mu}\tan{\beta}}{M_1^3}F_b\left(\frac{{m}^2_{\tilde \mu_L}}{M_1^2},\frac{{m^2_{\tilde \mu_R}}}{M_1^2}\right), \\
a_\mu({\tilde B}-{\tilde H},{{\tilde \mu}_L})&=&\frac{g_Y^2}{{16{\pi}}^2}\frac{{m_\mu^2{}M_1}{\mu}\tan{\beta}}{{m^4_{\tilde \mu_L}}}F_b\left(\frac{M_1^2}{m_{\tilde \mu_L}^2},\frac{\mu^2}{m_{\tilde \mu_L}^2}\right), \\
a_\mu({\tilde W}-{\tilde H},{{\tilde \mu}_L})&=&-\frac{g^2}{{16{\pi}}^2}\frac{{m_\mu^2{}M_2}{\mu}\tan{\beta}}{{m^4_{\tilde \mu_L}}}F_b\left(\frac{M_2^2}{m_{\tilde \mu_L}^2},\frac{\mu^2}{m_{\tilde \mu_L}^2}\right), \\
a_\mu({\tilde B}-{\tilde H},{{\tilde \mu}_R})&=&-\frac{g_Y^2}{{8{\pi}}^2}\frac{{m_\mu^2{}M_1}{\mu}\tan{\beta}}{{m^4_{\tilde \mu_R}}}F_b\left(\frac{M_1^2}{m_{\tilde \mu_R}^2},\frac{\mu^2}{m_{\tilde \mu_R}^2}\right),
\end{eqnarray}
where $g$ and $g_Y$ are the coupling constants of the SU(2) and U(1) gauge groups; $m_{\tilde \nu}$ and $m_{\tilde \mu}$ are the mass of the sneutrino($\tilde \nu$) and the smuon($\tilde \mu$), respectively;  $\mu$ is the Higgsino mass parameter;  $\tan\beta$ is the ratio of the two Higgs vacuum expectation values; $M_1$ and $M_2$ are the bino and the wino mass, respectively.
The functions $F_a(x,y)$ and $F_b (x,y)$ are defined as follows, 

\begin{figure}[h]
\begin{center}
\includegraphics[width=.4\textwidth, angle=0]{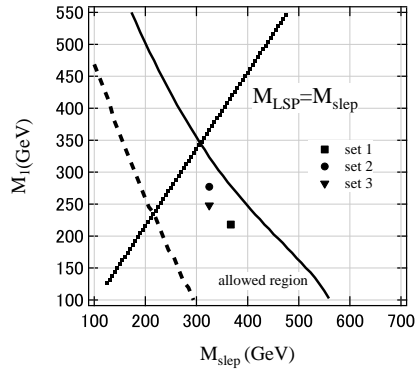}
\vspace{-10pt}
\caption{Allowed region from the muon $g$-$2$ constraint for $\tan\beta = 30$, $\mu = 600$GeV and $M_2=2M_1$.
The dotted line corresponds to $M_{LSP}=M_{slep}$. The solid and the dashed line correspond to  $\Delta a_{\mu}=1.76 \times{10}^{-9}$ and $\Delta a_{\mu}=4.00 \times{10}^{-9}$, respectively.}
\end{center}
\label{fig1}
\end{figure}

\begin{figure}[!]
\begin{center}
\hspace{-55pt}
\includegraphics[width=.8\textwidth, angle=0]{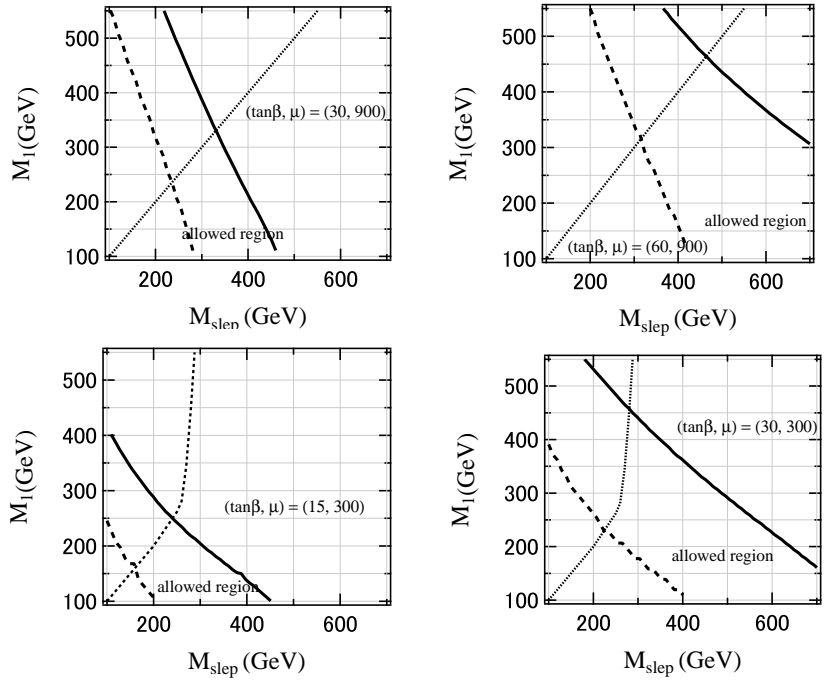}
\vspace{-55pt}
\caption{As in figure 1, but for a difference value of (tan$\beta$, $\mu$).}
\end{center}
\label{fig2}
\end{figure}

\begin{equation}
F_a(x,y) = - \frac{G_1(x)-G_1(y)}{x-y},\quad\quad F_b(x,y) = - \frac{G_2(x)-G_2(y)}{x-y},
\end{equation}
where 
\begin{eqnarray}
G_1(x)&=&\frac{1}{2(x-1)^3}((x-1)(x-3)+2 \log{x}), \\
G_2(x)&=&\frac{1}{2(x-1)^3}((x-1)(x+1)-2x \log{x}).
\end{eqnarray}

The allowed region obtained from the muon $g$-$2$ constraint in the ($M_{slep}$, $M_1$) plane  for  $\tan\beta = 30$ and $\mu = 600$GeV is shown in Figure 1, where $M_{slep}$ stands for $m_{\tilde \nu}$ $=$ $m_{\tilde \mu_L}$  $=$ $m_{\tilde \mu_R}+5$GeV  in eqs.(4) $\sim$  (8). 
The region between the solid curve and the dashed curve is allowed. 
The line corresponding to $M_{LSP}=M_{slep}$ is also plotted in Figure 1.
We consider the region under the line in which the lightest neutralino $\tilde \chi_1^{0}$ can be the stable LSP and a candidate for the DM. 
For reference, the points of three selected sets (Table 2, see sec.2.5) are shown by markers.
In Figure 2, we show the allowed region in  the ($M_{slep}, M_{1}$) plane for several other values of $\mu$ and tan$\beta$. While the allowed region changes with the value of  $\mu$ and $\tan\beta$,
we find that the upper limit of ($M_{slep}, M_{1}$)  is roughly (800,450) GeV.
The muon $g$-$2$ is the dominant constraint in our selection of the parameter sets and provides the credible experimental results which support ``natural SUSY".

\subsection{Observed value of the Higgs mass} \mbox{}
The measured mass of the Higgs boson is given by\cite{higgsatlus,higgscms} 
\begin{equation}
{m_h^{exp}} = 125.7{\pm}0.4\ \rm GeV. 
\end{equation}

\begin{figure}[h]
\begin{center}
\includegraphics[width=.6\textwidth, angle=0]{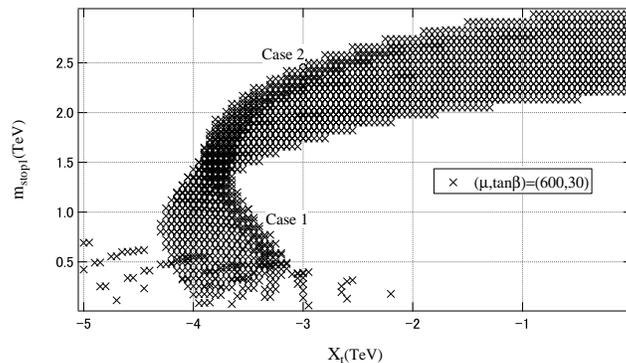}
\vspace{-15pt}
\caption{The contours on the  $(X_t, m_{\tilde t_1})$ plane which are consistent with the observed mass of the Higgs boson for  
$(\mu, \tan\beta)$ $=$ $(600$GeV$, 30)$.
For other MSSM parameters, we used the value of set 1 (Table 2, see sec.2.5).} 
\end{center}
\label{fig3}
\end{figure}

 In Figure 3, we show the region which is consistent with the observed Higgs mass in $(X_t, m_{\tilde t_1})$ plane, where 
\begin{eqnarray}
X_t&=&\frac{m_{\tilde t_2}^2 - m_{\tilde t_1}^2}{m_t}\sin{2\theta_t}=2(A_t -\mu \cot{\beta})
\end{eqnarray}
is the left-right mixing parameter of stops.
$A_t$ is a trilinear coupling which is one of the soft SUSY breaking parameters.
There are two cases which explain the observed Higgs mass: 
 (Case 1) the lighter stop mass $m_{\tilde t_1}$ $\nle$ $1.5$TeV and 
 $X_t$ $=$ $-(5 \sim 2)$TeV or  
 (Case 2) $m_{\tilde t_1}$ $>$ $1.5$TeV and $X_t$ $\nge$ $-4$TeV.
 Note that the Cases 1 and 2 correspond to the light and the heavy stop scenario, respectively.

\subsection{The DM abundance, $b \rightarrow s \gamma$ and $B_s \rightarrow \mu^+ \mu^-$} 

The detailed analysis of the fluctuation in the cosmic background radiation gives a severe constraint on the DM abundance\cite{p} by PLANCK observation.
\begin{equation}
{\Omega}{h^2}=0.1198{\pm}0.0026.
\end{equation}
Naively thinking, the higgsino-like LSP with  mass $\mathcal{O}(10^2)$GeV can be a good candidate for the DM, because their abundance becomes $\mathcal{O}(10^{-1})$\cite{dmmssm1}. 
However,  it is difficult to satisfy simultaneously the condition given by the recent measurements of the Higgs mass and the condition given by the branching ratios\cite{btosg},  
\begin{eqnarray}
&&{\rm Br}\left(B \to X_s \gamma \right)=\left(3.43\pm 0.21 \pm 0.07 \right)\times 10^{-4}, \\
&&{\rm Br}\left(B_s \to \mu^+\mu^- \right)=\left(3\pm 1 \right)\times 10^{-9}.
\end{eqnarray}
As a typical example of $A_t$ dependence of Br$(b\to s \gamma)$ and  $m_h$ are shown in Figure 4. 

\begin{figure}[h]
\begin{center}
\vspace{+75pt}
\hspace{-55pt}
\includegraphics[width=0.65\textwidth, angle=0]{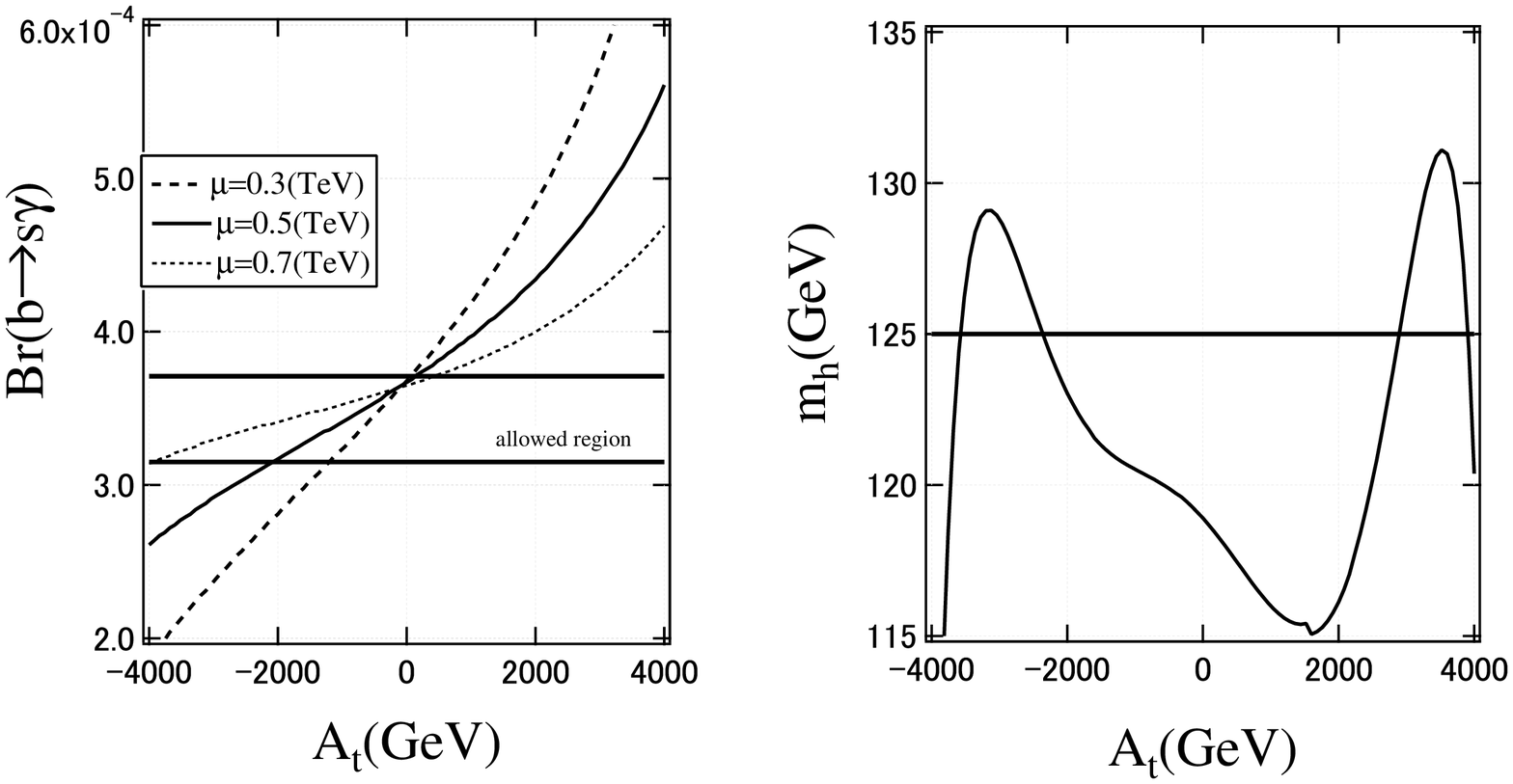}
\vspace{-75pt}
\caption{The $A_t$ dependence of Br$(b\to s \gamma)$ (left) and  $m_h$  (right). 
For other MSSM parameters, we used the value of set 1 (Table 2, see sec.2.5).}
\end{center}
\label{fig4}
\end{figure}

We find that the allowed range, eq.(15) for Br$(b\to s\gamma)$ and observed Higgs mass eq.(12) can be achieved simultaneously only when $A_t\nle-2.5 $TeV and $\mu$ $\nge$ $0.5$ TeV  in Figure 4. 
Here we recall the upper bound for the bino mass $M_1 \nle 0.45$TeV (obtained from the muon $g$-$2$ constraint) and it must be smaller than the higgsino mass $\mu$. 
Namely, the lightest neutralino $\widetilde{\chi}_1^0$ should be almost bino with small wino and higgsino components. 

\begin{figure}[!]
\begin{center}
\vspace{+55pt}
\hspace{-55pt}
\includegraphics[width=0.65\textwidth, angle=0]{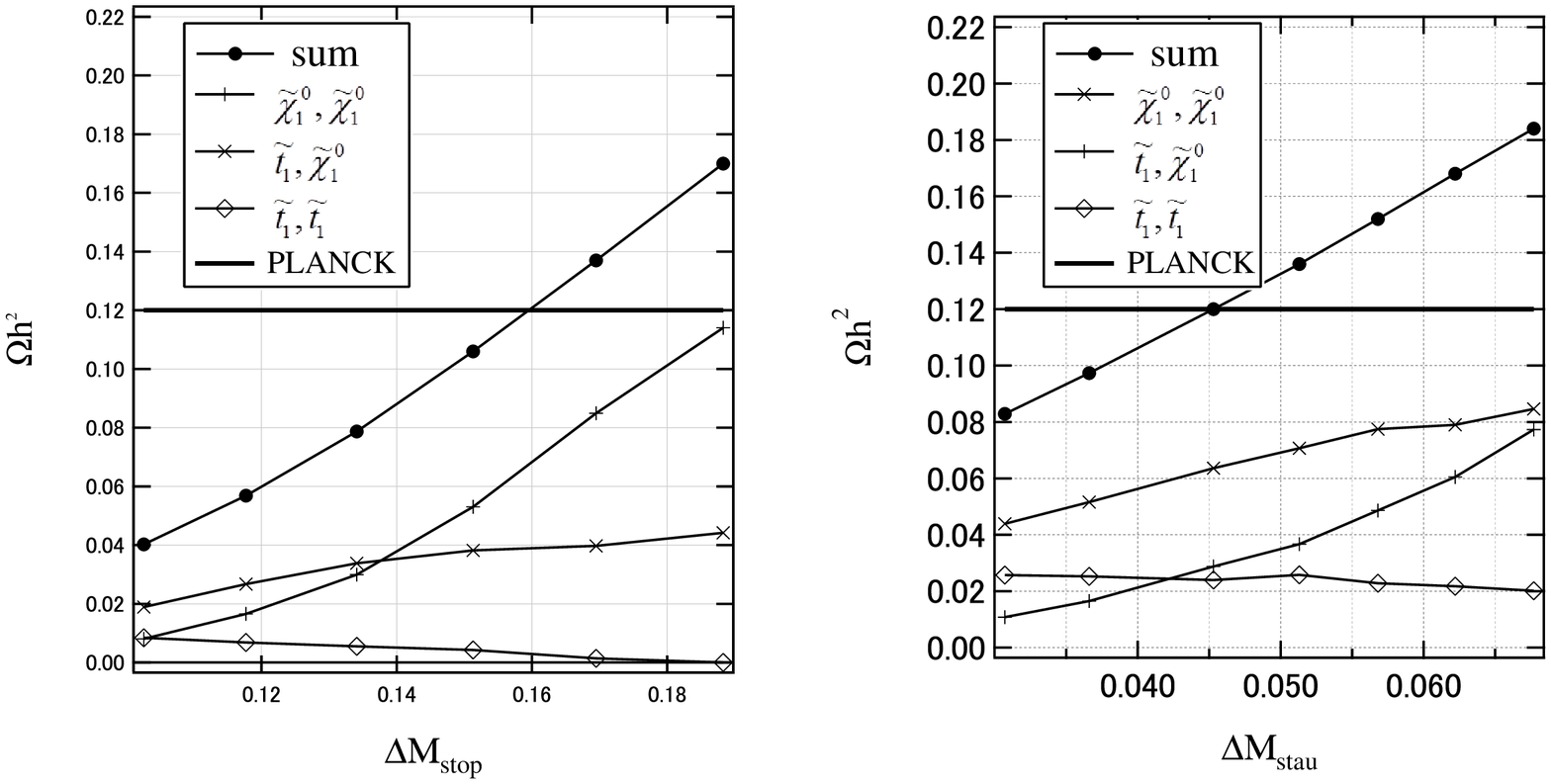}
\vspace{-75pt}
\caption{The dependence of the DM abundance on $\Delta$$M_{stop}$ ($m_{\tilde{t}_1}$) in the left figure, and on $\Delta$$M_{stau}$ ($m_{\tilde{\tau}_1}$) in the right figure. 
The measured abundance obtained from the PLANCK satellite is also shown. 
For other MSSM parameters, we used the value of set 3 and set 1 for the left and  right figure, respectively (Table 2, see sec.2.5). }
\end{center}
\label{fig5}
\end{figure}
 
 To meet the DM constraint eq.(14) with bino LSP, the co-annihilation is required in addition to the simple annihilation processes\cite{dmob}. 
In the co-annihilation, the next lightest SUSY particle (NLSP) plays an important role. 
Here, we consider two cases, the NLSP is the lighter stop $\widetilde{t}_1$ (stopCA) or the lighter stau $\widetilde{\tau}_1$ (stauCA). 
The DM abundance depends sensitively on the mass degeneracy $\Delta M_{stop}$ and $\Delta M_{stau}$, which are defined by 
\begin{equation}
\Delta M_{stop} = \frac{m_{\tilde{t}_1}-m_{\tilde \chi^{0}_{1}}}{m_{\tilde \chi^{0}_{1}}},
\qquad \Delta M_{stau} = \frac{m_{\tilde{\tau}_1}-m_{\tilde \chi^{0}_{1}}}{m_{\tilde \chi^{0}_{1}}}.
\end{equation}
In Figure 5, we show the dependence of the DM abundance on these parameters. 
The DM abundance is the sum of contributions from the simple annihilation of 
    $\tilde{\chi}^0_1 \tilde{\chi}^0_1$ and the co-annihilation of 
    $\tilde{t}_1 \tilde{\chi}^0_1$, $\tilde{t}_1 \tilde{t}_1$, 
    $\tilde{\tau}_1 \tilde{\chi}^0_1$ and $\tilde{\tau}_1 \tilde{\tau}_1$. Thus, the result of PLANCK observation eq.(14) becomes a severe constraint on  $m_{\tilde{t}_1}$, $m_{\tilde{\tau}_1}$ and $m_{\tilde{\chi}_{1}^{0}}$.

\subsection{The searches for sparticles at the LHC} 

Despite the systematic searches for the sparticles at the LHC, the evidence of the SUSY has not yet been reported. 
The lower limits on the masses of the sparticles have been updated. 
For example, the lower limits of the masses of the first- and second-generation squarks and gluino\cite{glsq}, 
\begin{equation}
m_{{\tilde q}},  m_{\tilde g}\nge 1.5  {\rm TeV}
\end{equation}
have been obtained by the analyses of the events with large missing transverse energies. 

Given the renormalization group equations (RGE) for the mass parameters in the MSSM assuming the GUT, it is theoretically natural to assume that the first- and second-generation squarks $\tilde q$ ($q=u, d, c, s$) and the gluino $\tilde g$ have larger masses  than the sleptons $\tilde l$, the charginos $\tilde \chi^{\pm}$ and the neutralinos $\tilde \chi^0$ due to the strong interaction. 
The lower mass limits of the chargino1 and the neutralino2 are
$m_{\tilde \chi_1^{\pm}, \tilde \chi_2^{0}}\nge 700$GeV 
for  $m_{\tilde \chi_1^{0}}\nle 400$GeV, 
when both the chargino and the neutralino decay only into $e$ ($\mu$) through the selectron (smuon)\cite{neutralinochargino}. 
However, if they also decay to $\tau$ through the stau, there is no limit on  
$m_{\tilde \chi_1^{\pm}, \tilde \chi_2^{0}}$
for  $m_{\tilde \chi_1^{0}}\nge 200$GeV. 

\begin{figure}[h]
\begin{center}
\includegraphics[width=.7\textwidth, angle=0]{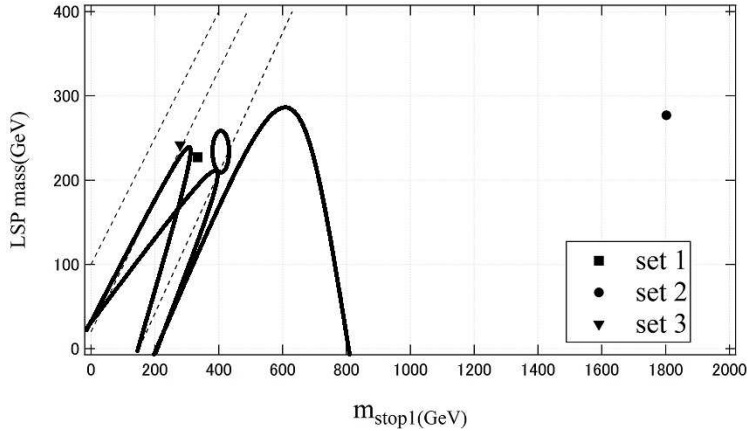}
\vspace{-15pt}
\caption{The excluded regions from the direct stop search at the LHC in $m_{\tilde{t}_1}$\textendash $m_{\tilde {\chi}^0_1}$ space\cite{directstop,cms}.} 
\end{center}
\label{fig6}
\end{figure}

The mass of the stop $\tilde{t}$ and the sbottom $\tilde{b}$ have different RGE evolution from the other squarks because  of their large Yukawa interactions in addition to the strong and electroweak interaction. 
The Yukawa interaction yields a negative contribution to the RGE of the mass parameters; 
thus, it is theoretically expected that $\tilde t$ and $\tilde b$ are lighter than $\tilde q$, $\tilde g$. 
Moreover, since the top quark mass is much larger than the other quark masses, there is a possibility that the difference between $m_{\tilde{t}_1}$ and the LSP mass $m_{\tilde \chi_1^0}$ is smaller than $m_t$. 
In this case, the stop has various possible decay modes $\tilde{t}_1$ $\to$ $b \tilde{\chi}_1^+$, 
$b W^+ \tilde{\chi}_1^0$, $b\ell^+{\tilde{\nu}_e}$, $b{\nu_e}\tilde{\ell^+}$, $c \tilde{\chi}_1^0$, $b q\bar{q} \tilde{\chi}_1^0$, $b \ell^+\nu \tilde{\chi}_1^0$, $u \tilde{\chi}_1^0$ depending on the MSSM parameters \cite{PhysRevD.36.724}. 
The excluded regions obtained from the analyses of the direct-stop searches at ATLAS and CMS are combined in Figure 6\cite{directstop,cms}.
As the $b$-quark is not so heavy, the sbottom $\tilde{b}$ can generally decay into $b+$LSP. 
The search strategies of  $\tilde{b}$ is, therefore, not so complicated compared with the $\tilde {t}_1$, and the limit $m_{\tilde b_1}\nge 800$GeV has been obtained\cite{sbottom}.

\subsection{Typical three MSSM parameter sets}

Based on discussion in the previous subsections, we classify  the MSSM parameter sets into the four categories shown in Table 1. Depending on the $\tilde {t}_1$ mass, two options, Case 1 and Case 2, are introduced in sec.2.2, while depending on whether NLSP is stau or stop, another two options, stauCA and stopCA, are introduced in sec.2.3.  Combining these two kinds of options, we have four categories as are shown in Table.1. Note that we have not chosen any set corresponding to Case 2 and StopCA. As has been mentioned in sec.2.1, we consider the muon $g$-$2$ constraint on the MSSM and 
adopt only  ``natural SUSY" sets with $m_{\tilde{\ell},\tilde{\chi}^0,\tilde{\chi}^{\pm}}$ $\nle$ $500$GeV.  In this case the LSP becomes so heavy $m_{\tilde{\chi}^0_1} \sim m_{\tilde{t}_1} \nge 1\rm{TeV}$ that the muon $g$-$2$ constraint is not satisfied.  As examples of Case1, we selected sets 1 and 3 with $m_{\tilde{t}_1}$ $\nle$ $350$GeV. As an example of Case2, we selected set 2 with $m_{\tilde{t}_1}$ $\simeq$ $1.8$TeV.
Thus we have selected three sets $1\sim 3$.

In Tables 2 and 3 the mass spectra and calculated important observables for the three sets are shown, respectively. 
Main decay modes of the major particles are shown in Table 4.
Note that the three sets are distinguished by the decay modes of the lighter stop. 
In set 1, $\tilde{t}_1$ mainly decays into $bW\tilde \chi^0_1$ because the mass difference between the $\tilde {t}_1$ and $\tilde \chi^{0}_{1}$ is larger than $m_W$. 
In set 2, $\tilde t_1$ decays into the $t\tilde g$ since $m_{\tilde t_1}$ is large enough. In set 3, $\tilde t_1$ can mainly decay into $c\tilde \chi^{0}_{1}$ with BR($\tilde{t}_1 \to c\tilde \chi^{0}_{1}) \simeq 99 \%$.
The main decay modes of $\tilde{b}_1$, $\tilde \chi^{\pm}_{1}$ and $\tilde \chi^{0}_{2}$ 
are also different for the three sets. 
On the other hand, the $\tilde{g}$ decay signals distinguish only the light stop and the heavy stop scenarios(Case 1 and Case 2). 

\begin{table}[ht]
\caption{Typical MSSM parameter sets}
\label{table1}
\centering
\LARGE
\begin{tabular}{|c|c|c|}
\hline
 & Case 1 & Case 2 \\ 
\hline
stauCA & set 1 & set 2\\
\hline
stopCA & set 3 & --- \\
\hline
\end{tabular}
\end{table}
\begin{table}[ht]
\caption{Masses and MSSM parameters for three sets (masses in unit of GeV)}
\label{table3}
\centering
\scalebox{1}{
\begin{tabular}{|c|c|c|c||c|c|c|c||c|c|c|c|}
\hline
\multicolumn{4}{|c||}{set 1} & \multicolumn{4}{|c||}{set 2} & \multicolumn{4}{|c|}{set 3} \\
\hline
\hline
$\tilde\chi^+_1$ &  $\tilde\chi^+_2$ &  &  &  $\tilde\chi^+_1$ &   $\tilde\chi^+_2$ &   &   &  $\tilde\chi^+_1$ &   $\tilde\chi^+_2$ &   &   \\
\hline
419.9  & 620.5 &  & &508.1  & 636.8 &  &  &  467.5  & 626.7 &  &\\
\hline
\hline
   $\tilde\chi^0_1$ &  $\tilde\chi^0_2$ &    $\tilde\chi^0_3$ & $\tilde\chi^0_4$& $\tilde\chi^0_1$ &  $\tilde\chi^0_2$ &    $\tilde\chi^0_3$ & $\tilde\chi^0_4$ & $\tilde\chi^0_1$ &  $\tilde\chi^0_2$ &    $\tilde\chi^0_3$ & $\tilde\chi^0_4$ \\
\hline
218.4 & 420.0 &  603.7 & 620.2 &    277.9 & 508.5 &  603.4 & 637.1 & 242.8 & 467.6 &  603.6 & 626.7 \\
\hline 
\hline
$\tilde \ell_1$ &  $\tilde \ell_2$ &  $\tilde\nu_\ell$ & & $\tilde \ell_1$ &  $\tilde \ell_2$  &  $\tilde\nu_\ell$ & &  $\tilde \ell_1$ &  $\tilde \ell_2$ &    $\tilde\nu_\ell$ &  \\
\hline
352.5 & 358.0 &  349.4 & & 317.8 & 323.3 &  313.8 & & 322.8 & 328.3 &  318.9 & \\
\hline
\hline
$\tilde\tau_1$ &  $\tilde\tau_2$ &    $\tilde\nu_\tau$ &  & $\tilde\tau_1$ &  $\tilde\tau_2$ &    $\tilde\nu_\tau$ &  & $\tilde\tau_1$ &  $\tilde\tau_2$ &    $\tilde\nu_\tau$ &  \\
\hline
228.4 & 336.3 & 277.9 &  &  283.9 & 377.1 &  327.4 & & 320.1 & 405.3 &  359.6 & \\
\hline
\hline
$\tilde u_1$ & $\tilde u_2$ & $\tilde d_1$ & $\tilde d_2$ &$\tilde u_1$ & $\tilde u_2$ & $\tilde d_1$ & $\tilde d_2$ &$\tilde u_1$ & $\tilde u_2$ & $\tilde d_1$ & $\tilde d_2$ \\
\hline
1719 & 1739 & 1740 & 1740& 1720 & 1739 & 1740 & 1741 & 1720 & 1739 & 1740 & 1741 \\
\hline
\hline
 $\tilde t_1 $ & $\tilde t_2$ & $\tilde b_1$ & $\tilde b_2$ & $\tilde t_1 $ & $\tilde t_2$ & $\tilde b_1$ & $\tilde b_2$ & $\tilde t_1 $ & $\tilde t_2$ & $\tilde b_1$ & $\tilde b_2$ \\
\hline
344.0 & 2078 & 899.9 & 2060.9& 1802 & 2244 & 1998 & 2063 & 279.6 & 2078 & 800.0 & 2061 \\
\hline
\hline
 $\theta_\tau$ &   $\theta_b$  &  $\theta_t$   &  & $\theta_\tau$ &   $\theta_b$  &  $\theta_t$   &  & $\theta_\tau$ &   $\theta_b$  &  $\theta_t$   & \\
\hline
0.7970  &   1.556  &  1.4502 & &  0.8150  &  1.376  &  0.8533  & & 0.8175  &   1.557  &  1.456  & \\
\hline
\hline
 $M_1$ &   $M_2$  &  $M_3$   & &  $M_1$ &   $M_2$  &  $M_3$   & &  $M_1$ &   $M_2$  &  $M_3$   & \\ 
\hline
220.0  &   435.0  &  2000   & & 280.0  &   540.0  &  1500   & & 244.5  &   489.0  &  2000   &  \\
\hline
\hline
\multicolumn{4}{|c||} {$\mu$=600,~~~$\tan\beta$=30} & \multicolumn{4}{|c||} {$\mu$=600,~~~$\tan\beta$=30} & \multicolumn{4}{|c|} {$\mu$=600,~~~$\tan\beta$=30} \\
\hline

\end{tabular}
}
\end{table}

\begin{table}[ht]
\caption{Important observables for three sets}
\label{table2}
\centering
\scalebox{0.9}{
\begin{tabular}{|c|c|c|c|c|c|}
\hline
  & $m_h$(GeV) & $\Delta a_\mu$($\times 10^{-9})$ &  $\Omega$$h^2$ & Br$(b\rightarrow$s$\gamma$)($\times 10^{-4}$) & Br($B_s \rightarrow \mu^+\mu^-$)($\times 10^{-9}$)   \\ 
\hline
 set 1  & 125.18 & 2.00  & 0.120  & 3.12  & 3.99   \\
\hline
 set 2  & 125.74  & 2.11  & 0.120 &3.11 &3.65 \\
\hline
set 3   &125.97 & 2.24 & 0.120 & 3.01  & 4.00  \\
\hline
\end{tabular}
}
\end{table}

\begin{table}[h]
\caption{Main decay modes of the major MSSM particles.}
\label{table4}
\centering
\scalebox{1.0}{
\begin{tabular}{|c|c|c|c|c|c|c|c|}
\hline
 &  $\tilde t_1 $ & $\tilde b_1$ & $\tilde g$ &  $\tilde\chi^+_1$ & $\tilde\chi^0_2$ & $\tilde\ell_{1,2}$   \\ 
\hline
  set 1 & $bW \tilde\chi^0_1$, $b \nu_{\tau} \tilde \tau_1$ & $t \tilde\chi^-_1$,  $b \tilde\chi^0_{3,4}$ & $t \tilde t_1 $, $b \tilde b_1 $  & $\tau^+ \tilde\nu_\tau$, $b$$\tilde t_1 $  & $\tau \tilde\tau_1$, $\bar\nu_\tau \tilde\nu_\tau$ & $\ell$$\tilde\chi^0_1$ \\
\hline
 set 2 &  $t$$\tilde g$, $b$$\tilde\chi^+_1$  & $b$$\tilde g$  & $q$$\bar q$$\tilde\chi^0_1$, $q$$\bar q$$\tilde\chi^\pm_1$  & $\ell^+$$\tilde\nu_\tau$,$\tau^+$$\tilde\nu_\tau$ &$\ell$$\tilde \ell_2$, $\tau$$\tilde\tau_1$, $\nu$$\tilde\nu$ &$\ell$$\tilde\chi^0_1$ \\
\hline
set 3 & $c$$\tilde\chi^0_1$ &$b$$\tilde\chi^0_1$, $t$$\tilde\chi^-_1$ & $t$$\tilde t_1 $, $b$$\tilde b_1 $ &  $b$$\tilde t_1 $, $\ell^+$$\tilde\nu_l$ & $t$$\tilde t_1$  & $\ell$$\tilde\chi^0_1$ \\
\hline
\end{tabular}
}
\end{table}

\clearpage
\section{Numerical results for the $e^-e^+\to\tau^-\tau^+, t\bar{t}, b\bar{b}$}
\subsection{The $\tt GRACE$ system and the calculation schemes} 

\ There are more than twice as many different types of particles in the MSSM as those in the SM ; therefore, there are various possible sparticle production processes in the collider  experiments. 
A large number of Feynman diagrams appearing in each production process
requires tedious and lengthy calculations in evaluating the cross sections.
 Accurate theoretical prediction requires an automated system to manage  such large scale computations.
$\tt GRACE$ system for the MSSM calculations\cite{grace1,grace2} has been developed in the KEK group (the Minami-tateya group) to meet the requirement. 
The $\tt GRACE$
system uses a renormalization prescription that imposes mass shell conditions on as many
particles as possible, while maintaining the gauge symmetry by setting the renormalization
conditions appropriately\cite{grace2}.
In the $\tt GRACE$ system for the SM, the usual 'tHooft-Feynman linear gauge 
  condition is generalized to a more general non-linear gauge (NLG) 
  that involves five extra parameters\cite{B, grace3}. 
  We extend it to the MSSM formalism by adding the SUSY interactions 
  with seven NLG parameters\cite{grace2, Baro}. We can check the consistency of the gauge symmetry by verifying the independence of the physical results from the NLG parameters. 
We ascertain that the results of the automatic calculation are reliable by carrying out the following checks:
\begin{itemize}
\item{Electroweak(ELWK) non-linear gauge invariance check (NLG check)}
\item{Cancellation check of ultraviolet divergence (UV check)} 
\item{Cancellation check of infrared divergence (IR check)}
\item{Check of soft photon (gluon) cut-off energy independence ($k_c$ check)} 
\end{itemize}
 
Actually, the 1-loop differential cross sections (distributions) are separated into two parts, 
 \begin{equation}
 d\sigma_{\rm L\&S}^{\rm M, G}(k_c) \equiv d\sigma_{\rm virtual}^{\rm M, G}  + d\sigma_{\rm soft}^{\rm G},
 \end{equation}
where, M=(SM or MSSM), G=(ELWK or QCD), and each point is computed separately. The loop and the counter term contribution $d\sigma_{\rm virtual}^{\rm M, G}$ should be gauge invariant and the UV finite but IR divergent. 
We regularize the IR divergence by the fictitious photon (or gluon) mass $\lambda$, so both $d\sigma_{\rm virtual}^{\rm M, G}$ and the soft photon (or gluon) contribution 
$d\sigma_{\rm soft}^{\rm G}$ are $\lambda$ dependent. 
The $\lambda$ dependence is canceled in $d\sigma_{\rm L\&S}^{\rm M, G}$.
Finally, the $k_c$ independent 1-loop physical cross sections can be obtained by

 \begin{equation}
 d\sigma_{\rm 1loop}^{\rm M, G} \equiv d\sigma_{\rm tree} + d\sigma_{\rm L\&S}^{\rm M, G}(k_c)+ \iint_{k_c}{ {\frac{d\sigma_{\rm hard}^{\rm G}}{d\Omega dk}}d\Omega dk},
 \end{equation}
where $k$ and $\Omega$ are the energy and the solid angle of the photon (or gluon). Strictly speaking, the tree level $d\sigma_{\rm tree}$, the soft $d\sigma_{\rm soft}^{\rm G}$ and the hard photon (gluon) $d\sigma_{\rm hard}^{\rm G}$ contributions are different for the SM and the MSSM because their Higgs contributions are not identical. 
However, the difference is numerically negligible at least in the present processes $e^-e^+\to f\bar f$, 
so we omitted the superfix M from $d \sigma_{tree}$ etc.
In order to verify the signature of existence of the new physics, it is desirable to minimize the uncertainty coming from the numerical integration. 
Thus, 
we define the ratio of the differential cross sections\cite{hollik},
\begin{equation}
\delta_{\rm susy}^{\rm G}\equiv{\frac{d\sigma_{\rm 1loop}^{\rm MSSM, G}-d\sigma_{\rm 1loop}^{\rm SM, G}}{d\sigma_{\rm tree}}},
\end{equation}
which allows an accurate  evaluation of  the difference between  effects of the MSSM  and the SM at 1-loop level. 
Because the tree cross section and the hard photon (or gluon) conributions disappear in the numerator of (21), $\delta_{\rm susy}^{\rm G}$ can  be rewritten as
\begin{equation}
\delta_{\rm susy}^{\rm G}={\frac{d\sigma^{\rm MSSM,G}_{\rm L\&S}-d\sigma_{\rm L\&S}^{\rm SM, G}}{d\sigma_{\rm tree}}}.
\end{equation}
Finally, we define the correction ratio of the sum of the ELWK and the QCD contributions as
\begin{equation}
\delta_{\rm susy} \equiv \delta_{\rm susy}^{\rm ELWK}+\delta_{\rm susy}^{\rm QCD}.
\end{equation}

\subsection{Selection of the beam energies}
\begin{figure}[ht]
\begin{center}
\includegraphics[width=.40\textwidth, angle=0]{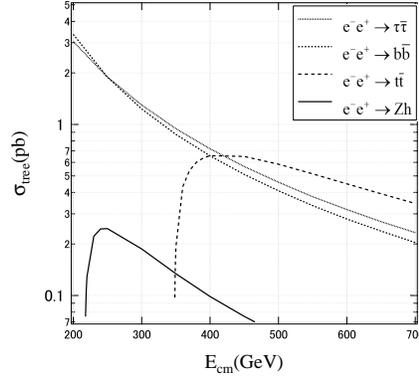}
\vspace{-15pt}
\caption{$E_{cm}$ dependence of the tree-level total cross section for $e^-e^+\to \tau^-\tau^+, b\bar b, t\bar t$ and $Zh$.}
\end{center}
\label{fig7}
\end{figure}
We show the center of mass energy ($E_{cm}$) dependence of the tree-level total cross sections for 
$e^-e^+ \to \tau^-\tau^+, b\bar b, t\bar t$ and $Zh$ in Figure 7.
One of the main purposes of the ILC project is the detail investigation of properties of the Higgs particle with many Higgs events; 
this will be realized at $E_{cm}=$ 250 GeV, which is almost the peak energy for $e^-e^+\to Zh$. 
Since the first stage of ILC experiments are planned at this $E_{cm}$, we choose to investigate the SUSY 1-loop effects in  $e^-e^+ \to \tau^-\tau^+, b\bar b$ at $E_{cm}=$ 250 GeV. The second stage of the ILC are planned at $E_{cm}=$ 500 GeV.
Therefore we investigate  $e^-e^+ \to t\bar t$ as well as $e^-e^+ \to \tau^-\tau^+, b\bar b$ at $E_{cm}=$ 500 GeV.

We find from Figure 7 that the tree level total cross section $\sigma(e^-e^+ \to \tau^-\tau^+, b\bar b)\simeq (2.0, 0.5)$pb for $E_{cm} = (250, 500)$GeV. 
In estimating the statistical errors of the cross sections, we assume the integrated luminosities $L = (250, 500)$ $\rm{fb^{-1}}$ at $E_{cm} = (250, 500)$GeV, which are planned values at the ILC project\cite{ilcd}. 
Therefore, for example, the statistical error of the total cross section is 
$\Delta\simeq (0.15,0.20)$\% for $E_{cm} = (250, 500)$GeV. 
As for the physical distributions, the estimated errors depend on the number of bins ($N_{\rm bin}$). In the following calculation, we take $N_{\rm bin} = 20$. 
\subsection{$e^-e^+ \to \tau^-\tau^+, b\bar b$ at $E_{cm} =$ 250 GeV} 
\begin{figure}[h]
\begin{center}
\includegraphics[width=.40\textwidth, angle=0]{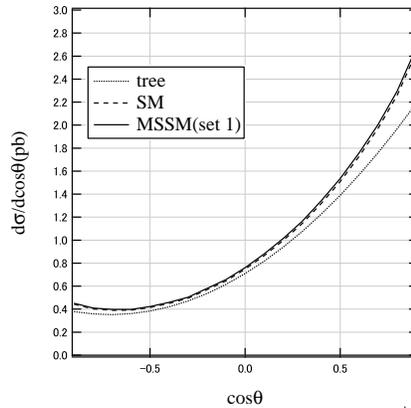}
\vspace{-15pt}
\caption{1-loop corrected angular distribution of $\tau^-$ in $e^-e^+ \to \tau^-\tau^+$ at $E_{cm}$ = 250 GeV. Dotted, dashed and solid line  correspond to the tree, SM 1-loop and MSSM (set 1) 1-loop level cross section, respectively.}
\end{center}
\label{fig8}
\end{figure}

\begin{figure}[h]
\begin{center}
\hspace{-55pt}
\includegraphics[width=.4\textwidth, angle=0]{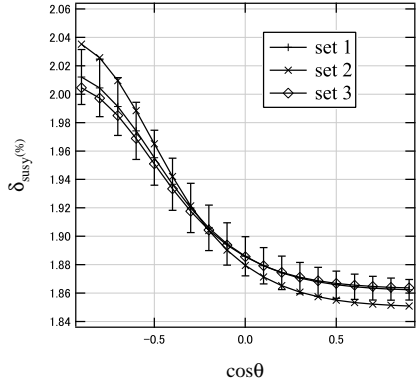}
\end{center}
\vspace{-15pt}
\caption{Correction ratios for three parameter sets in $e^-e^+\to\tau^-\tau^+$ at $E_{cm}$ = 250 GeV.}
\label{fig9}
\end{figure}
\begin{figure}[!]
\begin{center}
\vspace{+75pt}
\hspace{-55pt}
\includegraphics[width=.70\textwidth, angle=0]{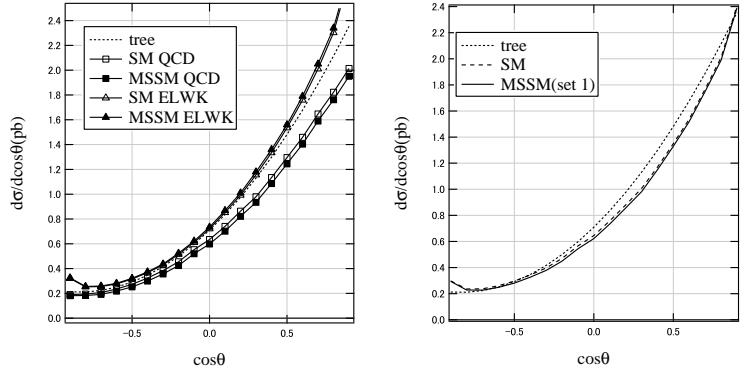}
\vspace{-85pt}
\caption{1-loop corrected angular distributions of $b$ in $e^-e^+\to b\bar b$ at $E_{cm}$ = 250 GeV. 
The left figure shows a detail of the QCD and ELWK corrections separately and in the right, the total corrections are shown.}
\end{center}
\label{fig10}
\end{figure}
We show the ELWK 1-loop corrected angular distribution of $\tau^-$ in $e^-e^+\to \tau^-\tau^+$ at $E_{cm}$ = 250 GeV in Figure 8.
The 1-loop correction is larger for the forward and the backward regions (e.g., $\sim 20\%$ at $|\cos\theta|\simeq 0.9$) than the central region (e.g., $\sim 6\%$ at $\cos\theta\simeq 0$). 
Since the measurements at the ILC are expected  to be accurate to within an error of a few percent,  the ELWK 1-loop corrections must be included in the theoretical prediction of the physical distributions. Since the main contribution of the ELWK correction is dominated by the SM contribution, the difference between the MSSM and the SM cross section is small. 
only the results of set1 are plotted in Figure 8 (set 2 and set 3 give essentially identical  results).

The values of $\delta_{\rm susy}$($=$$\delta_{\rm susy}^{\rm ELWK}$) for each parameter set are shown in Figure 9. 
For all the three sets, $\delta_{\rm susy}^{\rm ELWK}$ is $\sim 2\%$ 
and the statistical error for each bin is about $\pm 1\%$ over the entire region.
The dominant contributions  come from the Feynman diagrams with the vertex and box type loops, which depend on masses and couplings of $\tilde \ell_{1,2}$, $\tilde \nu_\ell$, $\tilde \chi_{i}^{0}$, $\tilde \chi^\pm_k$, $\tilde \tau_{1,2}$ and $\tilde \nu_\tau$.
If measurements at the ILC will be carried out within the 2\% accuracy, the deviation from the SM prediction for the three sets would be statistically verifiable. 
On the other hand, however, discrimination among the three parameter sets is substantially difficult.

We show the ELWK and QCD 1-loop corrected angular distributions of the $b$-quark in $e^-e^+\to b\bar b$ in Figure 10. 
At around cos $\theta$ $\simeq$ 0.9, the ELWK and QCD correction are about $+15\%$ and $-20\%$, respectively. 
The positive ELWK and the negative QCD contributions cancel each other over the entire region, so the magnitude of the MSSM 1-loop correction  is about $0.0 \pm 5.0\%$.

The values of $\delta_{\rm susy}^{\rm G}$ for $b \bar{b}$ production are shown in Figure 11. 
In the right figure, we find that $\delta_{\rm susy} \simeq (-5.0\sim -0.5)\%$ are almost equal among the three parameter sets. In the left figure, we plot $\delta_{\rm susy}^{\rm ELWK}$ and$\delta_{\rm susy}^{\rm QCD}$ for set 1 separately The ELWK correction is about $+2\%$, similar to $\tau^-\tau^+$ process, and the QCD correction is $(-7.0\sim -2.5)\%$.
Since the statistical error for each bin is $\pm (0.5\sim 1.0)\%$ over the entire region, the deviation from the SM in all parameter sets would be statistically verifiable with sufficiently high experimental accuracy.
We emphasize that evidence of the MSSM, which is manifested as a 5\% difference in bottom-pair production between the SM and the MSSM, might be confirmed during the early stage of experiments at the ILC. 
\begin{figure}[!]
\begin{center}
\vspace{+75pt}
\hspace{-55pt}
\includegraphics[width=.80\textwidth, angle=0]{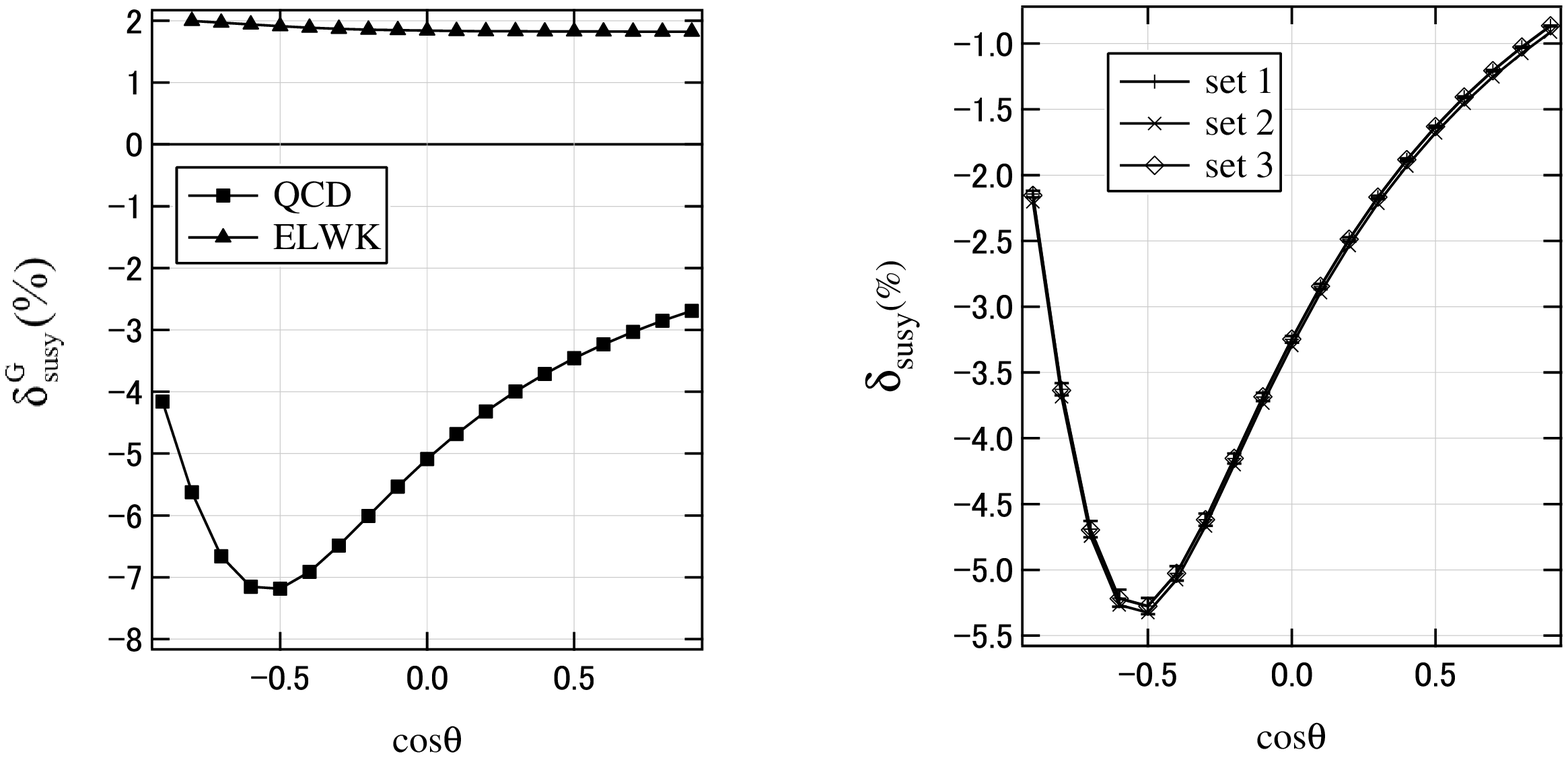}
\vspace{-95pt}
\caption{The correction ratio of QCD and ELWK for $e^-e^+\to b\bar b$ at $E_{cm} = 250$ GeV.
In the left, 1-loop QCD and ELWK contributions are shown separately(set 1), and in the right, the total corrections are shown.}
\end{center}
\label{fig11}
\end{figure}

\subsection{$e^-e^+ \to \tau^-\tau^+, b\bar{b}$ and $t\bar{t}$ at $E_{cm}$ = 500 GeV.}

\begin{figure}[h]
\begin{center}
\includegraphics[width=.40\textwidth, angle=0]{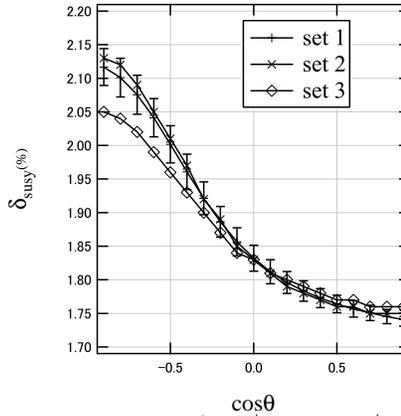}
\vspace{-15pt}
\caption{The correction ratio for  $e^+$$e^-$$\rightarrow$$\tau^-$$\tau^+$ at $E_{cm}$ = 500 GeV.}
\end{center}
\label{fig12}
\end{figure}

In Figure 12, the value of   $\delta_{\rm susy}$($=$$\delta_{\rm susy}^{\rm ELWK}$) for $e^-e^+\to\tau^-\tau^+$
are shown for  the three parameter sets.
The difference between set 1,2 and set 3 is larger than the statistical error in the backward direction. 
For example, the difference of $\delta_{\rm susy}$ between set 1 and set 3 becomes 0.04\% at (cos $ \theta$ $\simeq$ $-0.9$). 
There may be a possibility of distinguishing the set 3 from the others at $E_{cm} = 500$GeV. 
We don't show any plot for $b\bar b$ production at $E_{cm} = 500$GeV because we can easily expect to obtain similar results to those at $E_{cm} = 250$GeV. 
The angular dependences of $d\sigma/d\cos\theta$ and $\delta_{\rm susy}$ are almost the same as Figures 10 and 11, respectively. 
Of course, the values of $d\sigma/d\cos\theta$ are uniformly about one-fourth of those at $E_{cm} = 250$GeV.

In Figure 13, we show the 1-loop corrected angular distributions of the top quark in $e^-e^+\to t\bar t$. The QCD and the ELWK corrections are separately shown in the left figure. 
Both the QCD and the ELWK corrections are almost the same value $(0\sim 10\%)$ in the backward region. 
In the forward region, on the other hand, 
the QCD and the ELWK correction are $(10\sim 20\%)$ and $(-15\sim 0\%)$, respectively. 
Two corrections cancel in the forward region, while they are additive in the backward direction and the sum becomes as large as $30\%$.
  
The values of $\delta_{\rm susy}^{\rm G}$ for $t\bar t$ production are shown in Figure 14. 
In the left figure, we find that the ELWK correction is about $0.7\%$, while the QCD correction is $(-3.7\sim -2.5)\%$. 
Summing $\delta_{\rm susy}^{\rm ELWK}$ and $\delta_{\rm susy}^{\rm QCD}$, $\delta_{\rm susy} \simeq (-3.2\sim -1.9)\%$ for the three parameter sets, and 
the statistical error for each bin is $\pm (0.5\sim 1.0)\%$ over the entire region. 
The deviation from the SM for all parameter sets are  verifiable if the statistical error of the ILC will be within 3\%. Moreover, it should be emphasized that the differences among three parameter sets are larger than the statistical error. 
For example, the difference between set 2 and set 3 is as large as  0.2\% at $\cos\theta$ $\simeq$ $-0.4$. There is a possibility of distinguishing  set 3 from sets 1 and 2 at $E_{cm} = 500$GeV.
\begin{figure}[!]
\begin{center}
\vspace{+75pt}
\hspace{-55pt}
\includegraphics[width=.80\textwidth, angle=0]{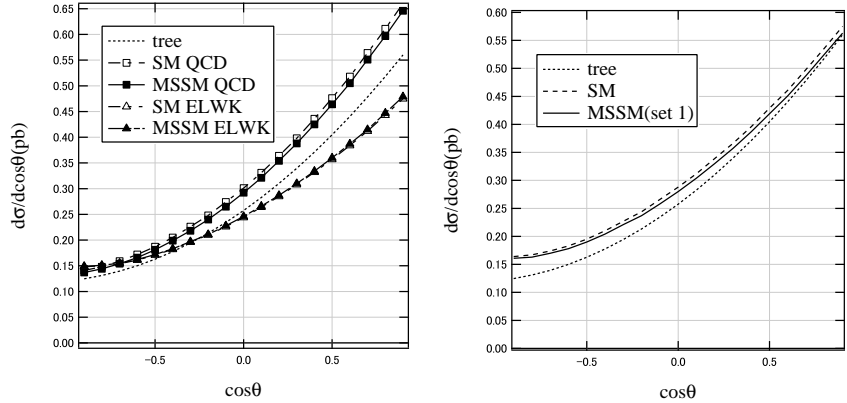}
\vspace{-85pt}
\caption{One-loop corrected angular distribution for $e^-e^+\to t\bar t$ at $E_{cm}$ = 500 GeV. The left figure shows a detail of the QCD and ELWK corrections separately and the right figure is the sum of the QCD and ELWK corrections.}
\end{center}
\label{fig13}
\end{figure}

\begin{figure}[h]
\vspace{+65pt}
\begin{center}
\hspace{-55pt}
\includegraphics[width=.80\textwidth, angle=0]{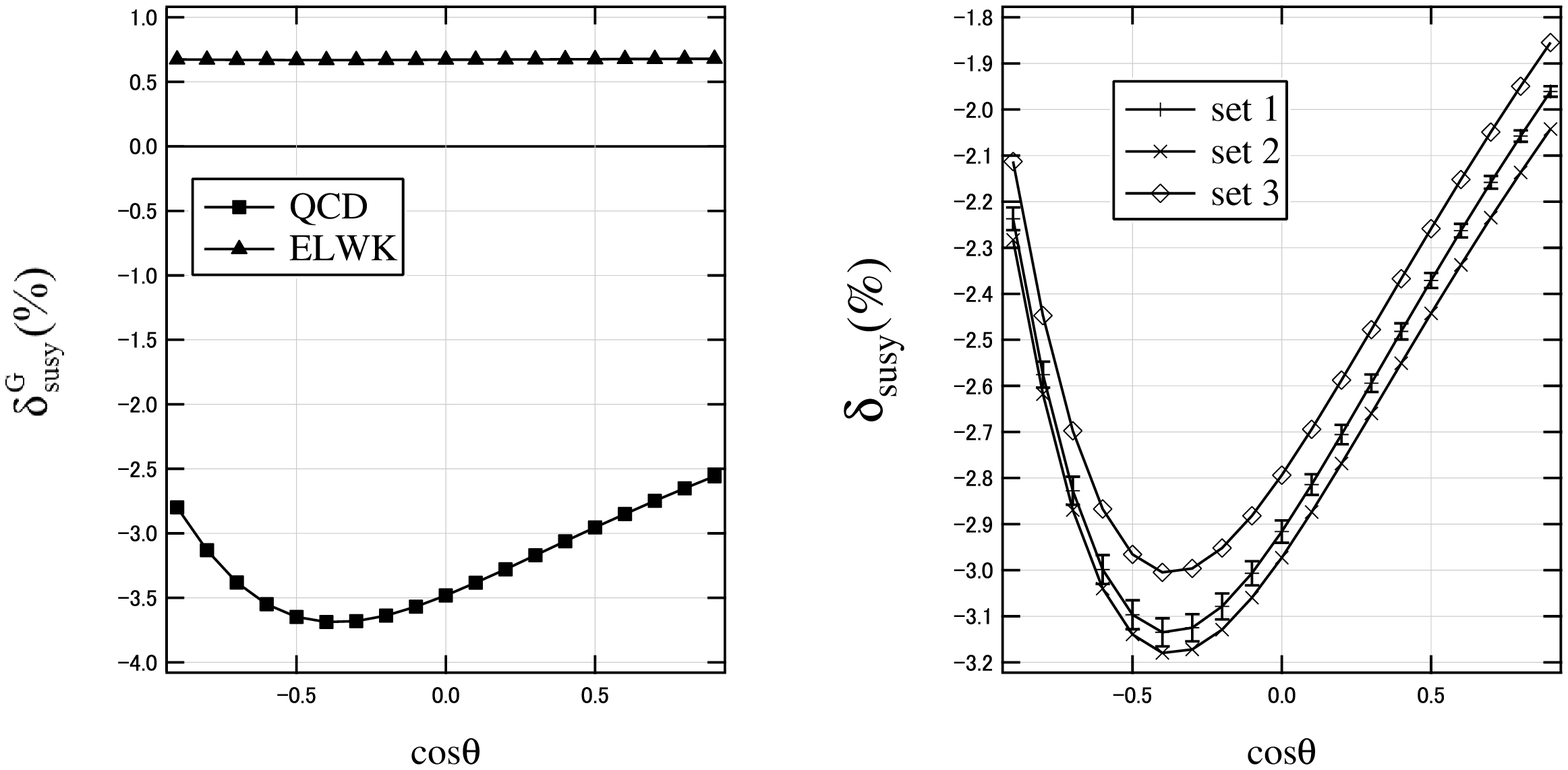}
\vspace{-95pt}
\caption{The correction ratio of QCD and ELWK for $e^-e^+\to t\bar t$ at $E_{cm}$ = 500 GeV.
In the left figure, 1-loop QCD and ELWK contributions are shown separately(set 1), and in the right figure, the total corrections are shown.}
\end{center}
\label{fig14}
\end{figure} 

\newpage
\section{Summary and conclusions}
We have obtained the MSSM sets for three scenarios which are consistent with the experimental results of  the Higgs mass, the muon $g$-$2$, the DM abundance, ${\rm{Br}}(b\to s\gamma)$, ${\rm{Br}}(B_s\to \mu^+\mu^-)$ and the direct search for the sparticles  at high energy colliders. 
For selected three typical parameter sets, the 1-loop level cross sections of $e^-e^+\to \tau^-\tau^+, b \bar b, t\bar t$ at $E_{cm}=250$ and $500$ GeV have been calculated by using $\tt GRACE/SUSY\mathchar`-loop$. 
If the sufficiently  accurate experiments are  realized at the ILC\cite{ilcd}, we will be able to verify the effects of the virtual sparticles in the selected parameter sets through the detailed measurements of the angular distributions of the third generation fermion-pair productions.
In particular, the effect of the MSSM for $b\bar b$ production at $E_{cm}=250$GeV will be expected to become about $5\%$ in the observable $\delta_{\rm susy}$. 
As for the discrimination of the parameter sets, $e^-e^+\to t\bar t$ at $E_{cm}=500$GeV seems to be the most promising.   
Although the scenario of the simple ``natural SUSY", for example, $m_{\tilde q, \tilde g}\nle 2$TeV (e.g., set 2), could be discovered at the LHC, other possible scenarios, for example, the case in which $m_{\tilde t} \sim m_{\tilde \chi_1^0}$ (set 3) or the case where $m_{\tilde \tau} \sim m_{\tilde \chi_1^0}$ (set 1), might be difficult to be explored at the LHC. 
We could confirm such scenarios if the above-mentioned processes in the ILC will be analyzed in detail. We emphasized that one loop effects of ``natural SUSY" in any credible scenarios would be visible at the ILC.
\vspace{5pt}

\end{document}